\documentclass[11pt,twoside]{article}
\usepackage{asp2010}
\resetcounters

\begin{document}

\title{How SFRIs vary with methods of sampling the IMF and duplicity}
\author{John J. Eldridge$^1$}
\affil{$^1$Institute of Astronomy, University of Cambridge, Madingley Road, Cambridge, CB3 0HA, England.}

\begin{abstract}
Using our new Binary Population and Spectral Synthesis (BPASS) code we
explore the affect of binary populations on the integrated spectra of
galaxies. We also explore the interplay of binary populations and a
varying maximum stellar mass. We compare our synthetic populations to
observations of H$\alpha$ emission from isolated clusters and
H$\alpha$ and FUV observations of galaxies. We find that observations
tend to favour a pure stochastic sampling of the initial mass function
although the evidence is not significant. We also find that binaries
make a stellar population less susceptible to the stochastic effects
of filling the IMF. Therefore making it more difficult to determine if
there is a variable maximum stellar mass.
\end{abstract}

\section{Introduction}

When observations are compared to the results from population
synthesis codes perfect agreement is rare. There are many uncertain
factors that can be tweaked to achieve a better agreement. The easiest
factor to adjust is the initial mass function (IMF). The detail that
is never adjusted, beyond altering the metallicity, is the stellar
evolution models.  This is probably because there are only so many
sets of available stellar models for input into population synthesis
and each synthesis code has its favourites. More over many stellar
evolution groups only provide their \textit{best} models rather than a
collection with varying mass-loss, mixing schemes, or binary
evolution.

Creating a full set of stellar models was an intensive process
requiring a great deal of user and computer time. Today with computer
time at a surplus and with recent developments leading to more
numerically stable evolution codes it is straight forward to create
many different sets of evolution models with varying input
physics. This allowed for large numbers of detailed models of binary
stars to be created \citep{EIT08}. Therefore we are now able to vary
the uncertain details of stellar evolution within our population
synthesis code together with other parameters such as the IMF.

Here we discuss how changing between single star to binary models
affects the star-formation rate indicators (SFRIs) of $H\alpha$ and
the UV luminosity. We also investigate the effect of assuming a
varying maximum stellar mass dependent on the star cluster mass on
these observables. First we outline our numerical method that uses the
recently developed \textit{Binary Population and Spectral Synthesis}
(BPASS) code \citep{es09}. Second we discuss how single stars and
binary star populations differ in their predictions for SFRIs when
applied to both star clusters and entire galaxies. Finally we
summarise our findings.

\section{Numerical Method}

\begin{figure}
\plotone{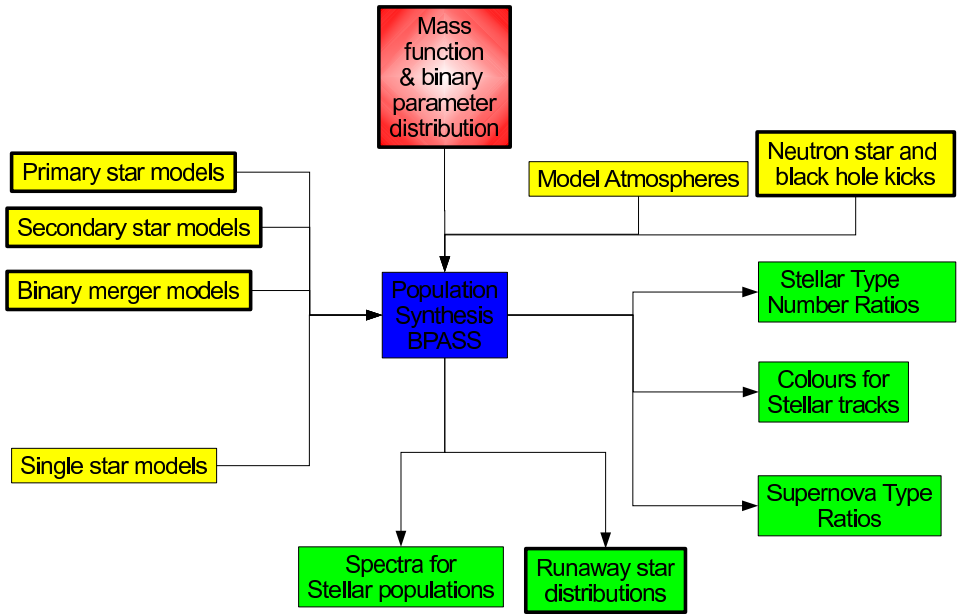}
\caption{Schematic diagram showing the inputs and outputs of
population synthesis codes. The boxes with thick borders indicate
those unique to BPASS.}
\label{fig:bpass}
\end{figure}

Recently we have developed a novel and unique code to produce
synthetic stellar populations that include binary stars
\citep{es09}. Figure \ref{fig:bpass} illustrates the inputs and
outputs that make this code unique. For example the binary models that
also require other new details such as the initial binary parameter
distribution and the effect of neutron star and black hole kicks to
determine the fate of the binaries after the first supernova. We also
obtain extra predicted outputs such as the velocity distribution of
runaway stars.

While similar codes exist, for example \citet{starburst99},
\citet{galev} and \citet{vanbev}.  BPASS has four important features
each of which set it apart from other codes and in combination make it
the state-of-the-art. First, and most important, is the inclusion of
binary evolution in modelling the stellar populations. The general
effect of binaries is to cause a population of stars to look bluer at
older ages than predicted by single-star models. Second is that
detailed stellar evolution models are used rather than an approximate
rapid population synthesis method. Third is the use of only
theoretical model spectra in our synthesis with as few empirical
inputs as possible to create a completely synthetic model to compare
with observations. Finally, the use of \textsc{Cloudy} \citep{cloudy}
to determine the nebular emission means we are modelling not only the
stars in detail but also the surrounding gas.

\begin{figure}
\plotone{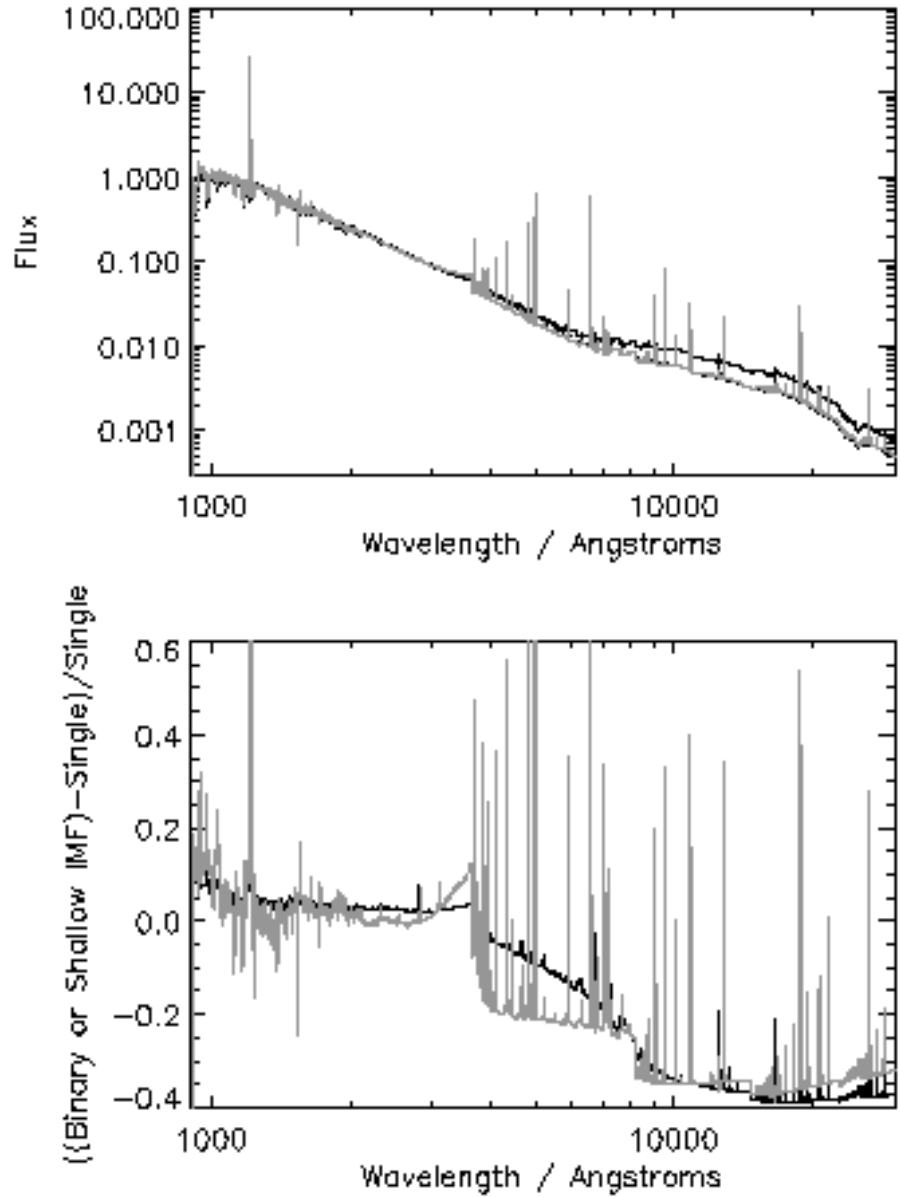}
\caption{Upper panel: the spectra of single star (solid black lines),
  binary (grey line) and shallow IMF (dashed black line)
  populations. The greatest difference is at longer wavelengths. Lower
  panel: fractional difference between single star spectrum and those
  of a single star population with shallow IMF (solid black line) and
  binary population (solid grey line).}
\label{fig:spec}
\end{figure}

To provide an example of the importance of binaries in stellar
populations we show in Figure \ref{fig:spec} how a binary stellar
population provides a similar spectrum to that of a single stellar
population with a shallow IMF of -1.35 (where Salpeter IMF is
-2.35). We see that the differences between a shallow IMF and binary
evolution are small, on the order of 10\%. The difference between them
to the normal Salpeter single star spectrum is up to 40\%. The main
identifying feature of the binary spectrum is that the spectrum has
more ionizing flux. This leads to the spectral shape being dominated
by the steps of the nebular continuum emission and the very strong
nebular emission lines, which may provide a method to distinguish if
we are observing a varying IMF or a binary dominated population.

BPASS and its comparison to various observations are discussed in
detail in \citet{EIT08} and \citet{es09}. Some results are available
for download at \texttt{http://www.bpass.org.uk}. The code is
versatile and adaptable to answer any question that requires knowledge
of stellar populations. Here we have altered the code to take account
of the stochastic nature of the filling of the IMF. We have created
multiple realisations of clusters and galaxies populating the stellar
and cluster IMFs at random rather than assuming a smooth IMF or
star-formation history. With the code we have tested two schemes of
filling the IMF. First, a \textit{pure stochastic sampling} (PSS),
where it is possible to have a $100M_{\odot}$ star in a $100M_{\odot}$
cluster (although it is very unlikely). The second scheme is that the
maximum mass of a star, $M_{\rm max}$, in a cluster, of mass $M_{\rm
  cl}$, is dependent on the total cluster mass. We calculate the
maximum star mass from
\begin{equation}
\log (M_{\rm max}/M_{\odot})= 2.56 \log (M_{\rm cl}/M_{\odot}) \big(3.82^{9.17}+(\log (M_{\rm cl}/M_{\odot}))^{9.17} \big)^{\frac{-1}{9.17}} - 0.38,
\end{equation}
which was derived by \citet{pflamm}. In this scheme in a
$100M_{\odot}$ cluster the maximum stellar mass is $9.1M_{\odot}$. We
refer to this as the \textit{variable maximum mass} (VMM) case. In
both cases we limit the most massive stars to $120M_{\odot}$.

We create the stellar population of a synthetic cluster by first
setting the maximum stellar mass in the cluster from one of the two
methods above. I.e. $120M_{\odot}$ in the PSS case or from the above
equation in the VMM case. Then we pick the masses of stars in the
cluster at random from the IMF. We use a Salpeter slope of -2.35 above
$0.5M_{\odot}$ and -1.3 below this mass to $0.1M_{\odot}$. We continue
to pick stars until the total mass of the stars is greater than our
target cluster mass. When the last star is added to the cluster we
examine the difference between the final and target cluster mass,
whether the difference is smaller with or without the last star. We
use the case with the smallest difference.

The situation for binary populations is slightly more
complicated. When we consider a binary population we assume that each
star above $5M_{\odot}$ has a binary companion and at random give the
star a mass ratio and separation. The mass ratio distribution is flat
over the range $0<M_2/M_1<1$ and a separation distribution is flat in
$\log a$ over the range of separations from 10 to $10^4 R_{\odot}$. We
include the mass of the secondary companion in the total mass of the
stellar population.

This is the case for a single cluster and can be repeated many times
to understand the range of possible clusters. We use many synthetic
clusters in Section \ref{iclust} below. When we simulate the stellar
population of an entire galaxy we do so by forming multiple stellar
clusters with masses determined from a cluster mass function. The
probability of a cluster mass is, $P(M_{\rm cl}) \propto M_{\rm
cl}^{-2}$. We take the minimum cluster mass to be $50M_{\odot}$ and a
maximum cluster mass of $10^6M_{\odot}$. A key difference in our
method to other such simulations is that rather than fixing the
star-formation history to some smooth function we give each cluster a
random age from over 100 Myrs. We then continue to create synthetic
clusters until the total mass of stars in the galaxy averaged over 100 Myrs
gives the required star-formation rate. For example to achieve a mean
star formation rate of $1 M_{\odot}/{\rm yr}$ requires $10^8M_{\odot}$
of stars. We find that the random star-formation history produces a
greater scatter into our predicted observables as seen below. Our
reasoning behind this is that nature forms clusters as distinct
components and there is no reason to expect different clusters to all
form at the same time.

Once we have our synthetic stellar populations of either single
clusters with a single age or a collection of clusters with a range of
ages we can predict various required observables. Here we concentrate
on two, the H$\alpha$ and UV fluxes that are both used as
SFRIs. H$\alpha$ is provided by stars more massive than $20M_{\odot}$,
while UV flux is produced by stars down to $3M_{\odot}$. Examining
these two indicators we have some idea regarding how the upper end of
the IMF is populated and which of PSS or VMM is in action in the
Universe.

\section{Results}

Theoretical models are limited in their use without comparison to
observations. Comparison both checks the model accuracy and enables us
to learn something about the physics in action in the Universe. Here
we compare our synthetic populations to observations of individual
star clusters and the emission from nearby galaxies.

\subsection{Individual Clusters}
\label{iclust}

\begin{figure}
\plottwo{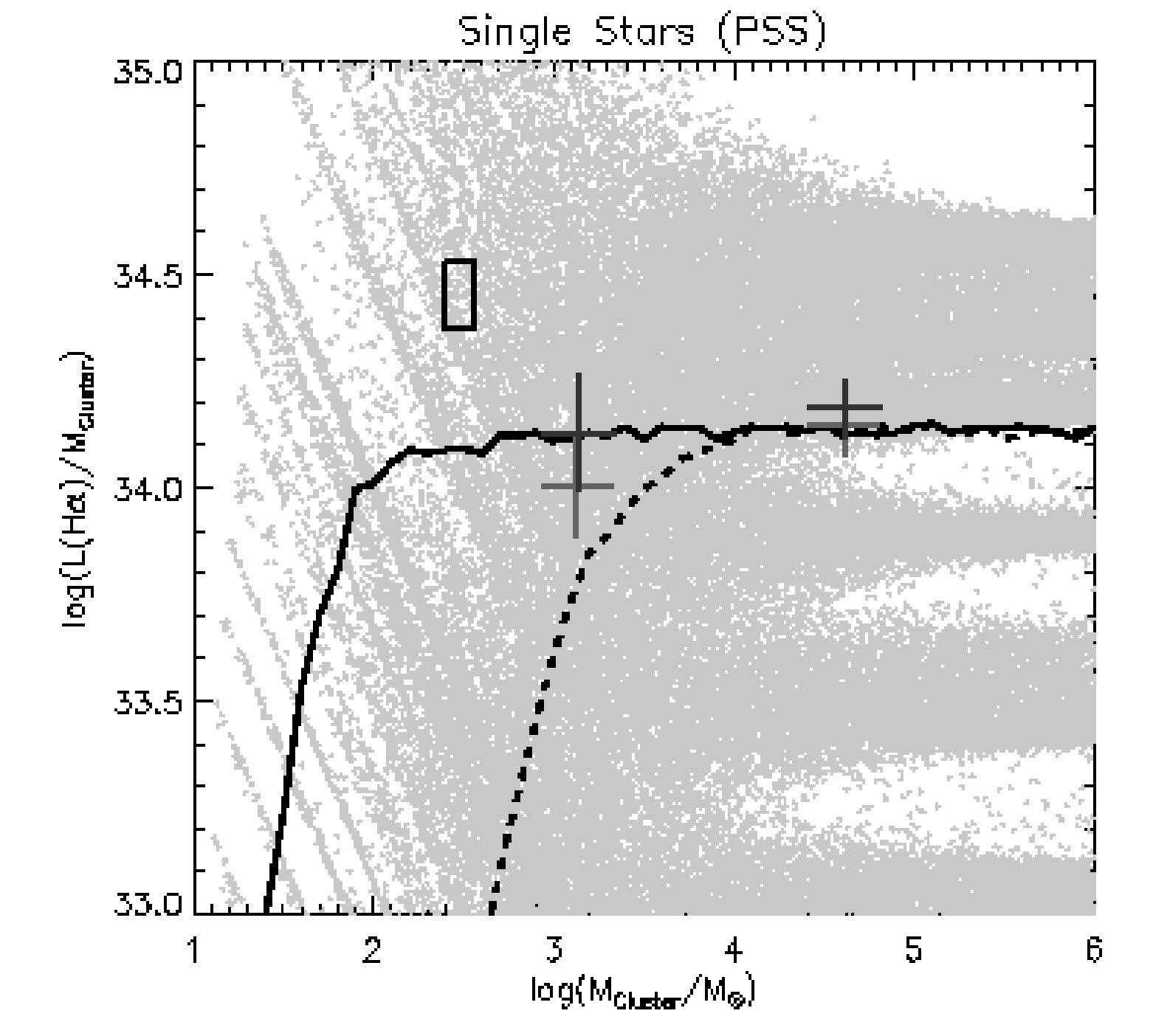}{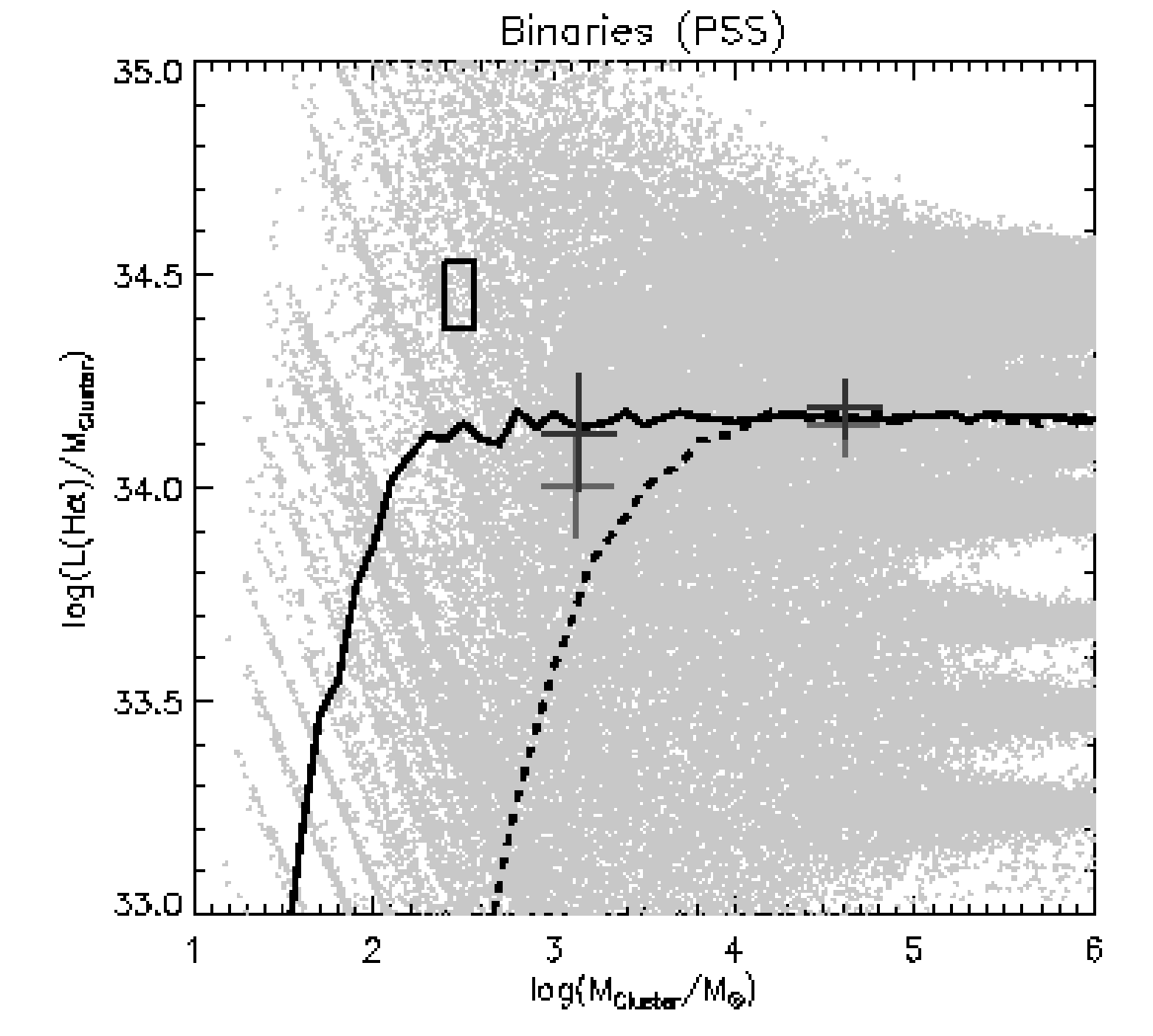}
\plottwo{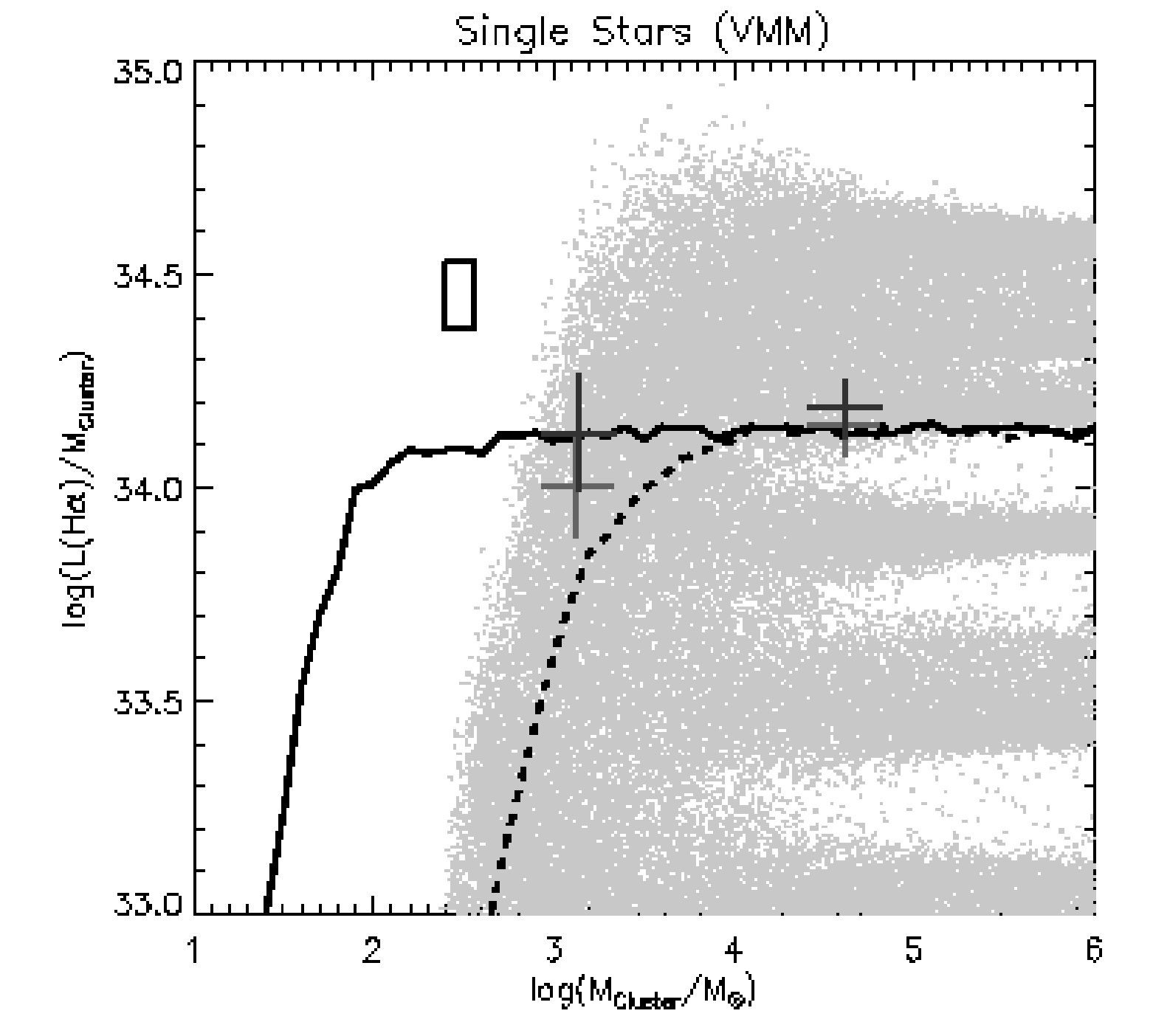}{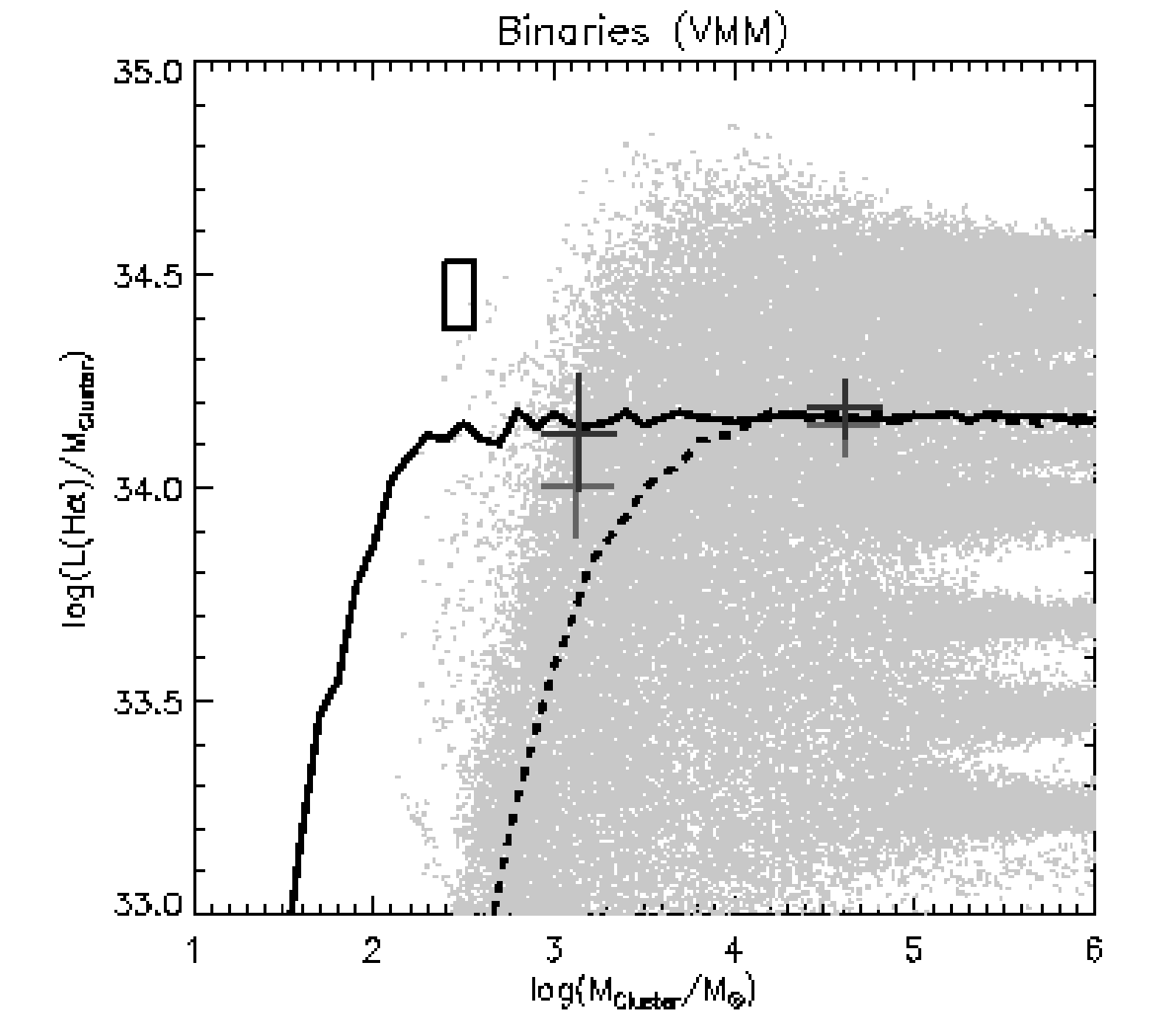}
\caption{Plot of H$\alpha$ flux per stellar mass of material in
  clusters. The points are individual realisations of different
  clusters from two methods of filling the IMF: PSS for upper two
  panels and VMM for the lower two panels. The lines show the mean
  for these two methods, solid black line for PSS and dashed black
  line for VMM. The grey crosses represent the observations of
  \citet{calzetti}: light grey including clusters not detected in
  H$\alpha$ and dark grey not including them. The black box represents
  the values derived from the Velorum cluster estimated from
  \citet{jeffries}, \citet{crowvel} and \citet{egamma}. The left
  panels are for clusters including single stars and the right panels are
  for binary populations. The linear features in the points of the
  upper panels are due to the limited resolution of the stellar
  model's initial masses.}
\label{fig1}
\end{figure}

Recently \citet{calzetti} have attempted to measure the H$\alpha$
emission from a number of young stellar clusters to determine
the mean H$\alpha$ emission per Solar mass of material in a
cluster. A number of clusters have been observed in H$\alpha$ that are
selected to have ages between 1 to 8Myrs from their broad-band
colours. In \citet{calzetti} the sample of clusters are then binned
into two mass ranges. The idea behind this is that if the sampling of
the IMF is by PSS then a hundred $1000M_{\odot}$ clusters should have
the same stellar population as one $10^5M_{\odot}$ cluster. This means
they should have the same H$\alpha$ flux per mass of stars. If there
is a VMM then we should expect fewer massive star and therefore less
H$\alpha$ per mass of stars. In Figure \ref{fig1} we compare synthetic
clusters to the observations of \citet{calzetti}. Making sure the
models and observations have the same mass range and age range. We also
include lines that represent the mean H$\alpha$ flux per mass for all
the synthetic clusters in the mass range.

The single star H$\alpha$ flux decreases more rapidly with age than
that for a binary population. Thus the single star population does
have lower minimum H$\alpha$ fluxes, by about 0.4 dex after
8Myrs. When comparing to the observations there are two pairs of points
from whether the clusters not detected in H$\alpha$ are included or
not. It is apparent that the points in general agree more closely to
PSS. Further evidence is supplied by the existence of the box in the
Figures representing the Velorum cluster that has a total mass of 250
to 360$M_{\odot}$. It includes two stars in a massive binary which
have initial mass of 35 and 30$M_{\odot}$. For such a cluster VMM
predicts a maximum stellar mass of $19M_{\odot}$. While we cannot rule
out there is a VMM this one example again points towards PSS being
more likely.

\subsection{11HUGS Galaxies}

\begin{figure}
\plottwo{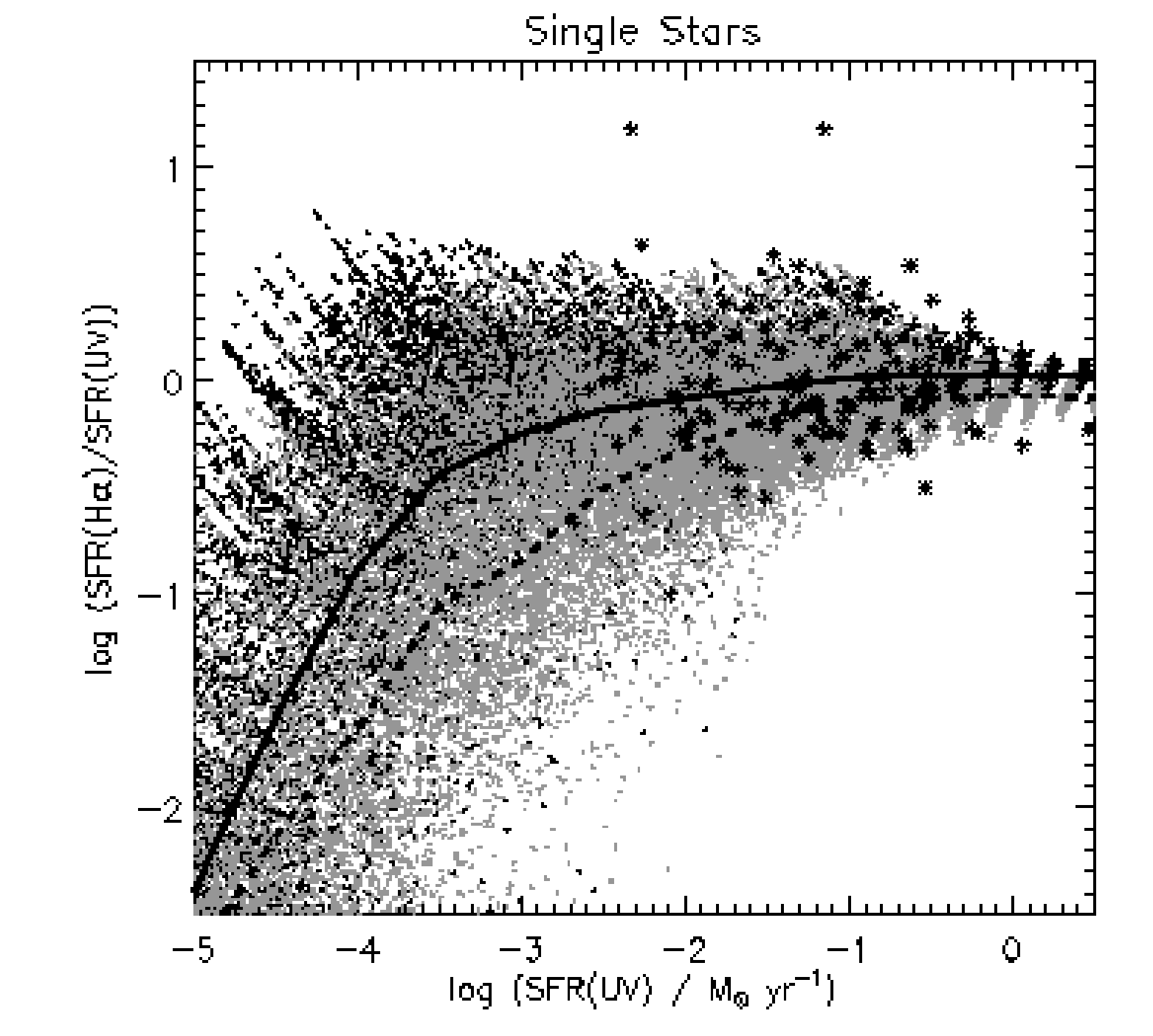}{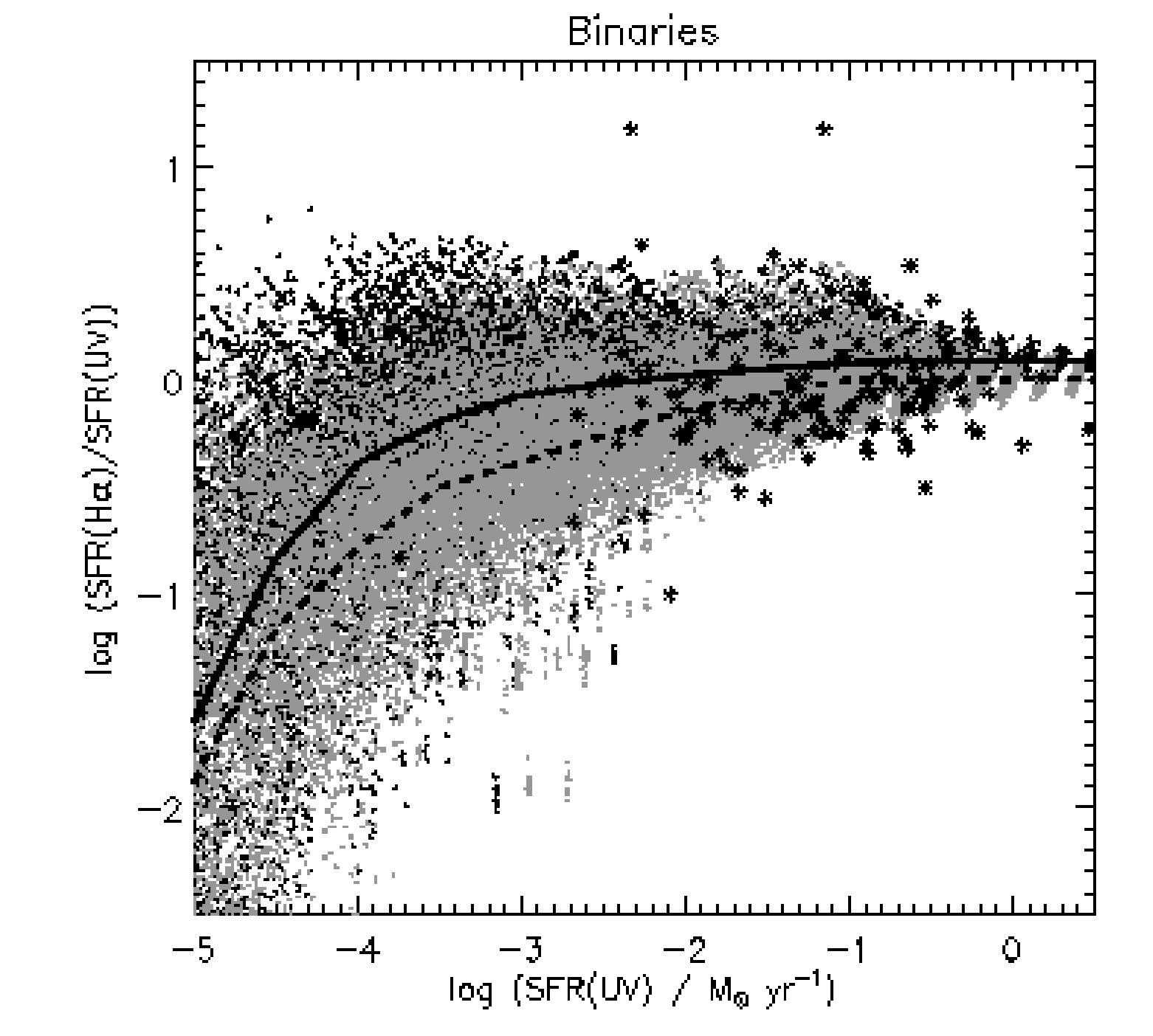}
\caption{The ratio of SFR measured by H$\alpha$ and UV fluxes verses
  the SFR from UV flux. The asterisks are the observations from
  \citet{lee} while the dots are individual realisations of synthetic
  galaxies. The lines indicate the mean ratios for PSS in the solid line
  and VMM for the dashed line. The left panel is for a single star
  population and the right panel for binary populations.}
\label{fig2}
\end{figure}

Another consequence of varying the method of filling the IMF is that
different SFRIs based on different mass ranges of stars will give
discrepant results at different galaxy-wide star-formation
rates. \citet{lee} estimated the SFR for a sample of galaxies from
both their H$\alpha$ and UV fluxes. These results are shown in Figure
\ref{fig2} again compared to synthetic stellar populations. In this
plot the synthetic galaxies assume constant star-formation over a
period of 100Myrs. The models predict that as the SFR decreases so to
does the ratio of SFRs measured by H$\alpha$ and UV. This is because
the H$\alpha$ flux is provided primarily by stars of over
$20M_{\odot}$ while stars down to $3M_{\odot}$ provide UV flux. At low
SFRs it becomes less likely to get the massive stars required but all
less likely for them to be formed recently enough to contribute to the
H$\alpha$ emission.

It has been suggested by \citet{jan} that the turn down of this ratio
as evidence for a VMM. However here we show that much of the observed
scatter can be explained by a stochastic star-formation
history. Furthermore this ratio is only an upper limit to the expected
ratio. There is growing evidence that the covering factor of young HII
regions is less than unity and therefore many ionizing photons might
escape galaxies and not produce H$\alpha$ emission. The UV flux is a
direct measure of the flux from the stars themselves so is a more
reliable measure of the actual mass of stars present.

In Figure \ref{fig2} if the single star and binary populations are
compared we see that there is little difference in the range of
scatter. The binary population ratio however does not decrease as
rapidly as the single star populations. This is because the effect of
binaries is to increase the number of hot and luminous stars and
therefore provide more H$\alpha$ over a longer timescale. Again we
should not expect such a rapid turn down as suggested by \citet{jan}.

Further work is required in simulating these galaxies but we see here
that to determine how the IMF is filled cannot be measured from the
H$\alpha$ to UV ratio unless a large number of very low star-formation
rate galaxies are observed. 

\section{Conclusions}

In population synthesis the IMF is an uncertain factor, so to is the
evolution of stars. Many of the results that are used to infer that
the IMF is variable may be explained away when binaries are included
in population synthesis.

The H$\alpha$ flux per Solar mass observed in samples of clusters
agrees with PSS of the IMF. However more observations are required to
rule out a VMM. Individual low-mass clusters with one or two massive
OB and WR stars are perhaps the most strict test to distinguish between PSS
and VMM and therefore more such clusters should be searched for.

The ratio of the H$\alpha$ and UV flux SFRs in galaxies has a
significant scatter that can be explained by the stochastic nature of
the star-formation history, rather than IMF variations or a VMM. The
ratio of SFRIs for stellar populations including binaries varies less
than that for populations of single star models only. This is because
binaries can experience mergers and mass-transfer and produce more
massive stars. However finding the low-star-formation rate galaxies
required to decide between PSS and VMM is difficult. Also measuring
the SFRIs accurately in such galaxies is difficult. Using a wider
range of SFRIs and observables beyond those presented here is a more
promising avenue to take to discriminate between PSS and VMM. This
requires further development of BPASS to include more physical
processes in greater detail.

\acknowledgements JJE would like to thank Benjamin Johnson, Monica
Relano, Dan Weisz, Daniella Calzetti, Janice Lee and Rob Kennicutt for
useful discussion. He is supported by the Theory Rolling Grant at the
Institute of Astronomy, University of Cambridge.

\bibliography{eldridge_j}

\begin{thebibliography}{}
\expandafter\ifx\csname natexlab\endcsname\relax\def\natexlab#1{#1}\fi
\expandafter\ifx\csname url\endcsname\relax
  \def\url#1{\texttt{#1}}\fi
\expandafter\ifx\csname urlprefix\endcsname\relax\def\urlprefix{URL }\fi
\providecommand{\eprint}[2][]{\url{#2}}

\bibitem[{{Calzetti} et~al.(2010){Calzetti}, {Chandar}, {Lee}, {Elmegreen},
  {Kennicutt}, \& {Whitmore}}]{calzetti}
{Calzetti}, D., {Chandar}, R., {Lee}, J.~C., {Elmegreen}, B.~G., {Kennicutt},
  R.~C., \& {Whitmore}, B. 2010, ArXiv e-prints. \eprint{1007.3188}

\bibitem[{{De Marco} et~al.(2000){De Marco}, {Schmutz}, {Crowther}, {Hillier},
  {Dessart}, {de Koter}, \& {Schweickhardt}}]{crowvel}
{De Marco}, O., {Schmutz}, W., {Crowther}, P.~A., {Hillier}, D.~J., {Dessart},
  L., {de Koter}, A., \& {Schweickhardt}, J. 2000, \aap, 358, 187.
  \eprint{arXiv:astro-ph/0004081}

\bibitem[{{Eldridge}(2009)}]{egamma}
{Eldridge}, J.~J. 2009, \mnras, 400, L20. \eprint{0909.0504}

\bibitem[{{Eldridge} et~al.(2008){Eldridge}, {Izzard}, \& {Tout}}]{EIT08}
{Eldridge}, J.~J., {Izzard}, R.~G., \& {Tout}, C.~A. 2008, \mnras, 384, 1109.
  \eprint{0711.3079}

\bibitem[{{Eldridge} \& {Stanway}(2009)}]{es09}
{Eldridge}, J.~J., \& {Stanway}, E.~R. 2009, \mnras, 400, 1019.
  \eprint{0908.1386}

\bibitem[{{Ferland} et~al.(1998){Ferland}, {Korista}, {Verner}, {Ferguson},
  {Kingdon}, \& {Verner}}]{cloudy}
{Ferland}, G.~J., {Korista}, K.~T., {Verner}, D.~A., {Ferguson}, J.~W.,
  {Kingdon}, J.~B., \& {Verner}, E.~M. 1998, \pasp, 110, 761

\bibitem[{{Jeffries} et~al.(2009){Jeffries}, {Naylor}, {Walter}, {Pozzo}, \&
  {Devey}}]{jeffries}
{Jeffries}, R.~D., {Naylor}, T., {Walter}, F.~M., {Pozzo}, M.~P., \& {Devey},
  C.~R. 2009, \mnras, 393, 538. \eprint{0810.5320}

\bibitem[{{Kotulla} et~al.(2009){Kotulla}, {Fritze}, {Weilbacher}, \&
  {Anders}}]{galev}
{Kotulla}, R., {Fritze}, U., {Weilbacher}, P., \& {Anders}, P. 2009, \mnras,
  396, 462. \eprint{0903.0378}

\bibitem[{{Lee} et~al.(2009){Lee}, {Gil de Paz}, {Tremonti}, {Kennicutt},
  {Salim}, {Bothwell}, {Calzetti}, {Dalcanton}, {Dale}, {Engelbracht}, {Funes},
  {Johnson}, {Sakai}, {Skillman}, {van Zee}, {Walter}, \& {Weisz}}]{lee}
{Lee}, J.~C., {Gil de Paz}, A., {Tremonti}, C., {Kennicutt}, R.~C., {Salim},
  S., {Bothwell}, M., {Calzetti}, D., {Dalcanton}, J., {Dale}, D.,
  {Engelbracht}, C., {Funes}, S.~J.~J.~G., {Johnson}, B., {Sakai}, S.,
  {Skillman}, E., {van Zee}, L., {Walter}, F., \& {Weisz}, D. 2009, \apj, 706,
  599. \eprint{0909.5205}

\bibitem[{{Leitherer} et~al.(1999){Leitherer}, {Schaerer}, {Goldader},
  {Gonz{\'a}lez Delgado}, {Robert}, {Kune}, {de Mello}, {Devost}, \&
  {Heckman}}]{starburst99}
{Leitherer}, C., {Schaerer}, D., {Goldader}, J.~D., {Gonz{\'a}lez Delgado},
  R.~M., {Robert}, C., {Kune}, D.~F., {de Mello}, D.~F., {Devost}, D., \&
  {Heckman}, T.~M. 1999, \apjs, 123, 3. \eprint{arXiv:astro-ph/9902334}

\bibitem[{{Pflamm-Altenburg} et~al.(2007){Pflamm-Altenburg}, {Weidner}, \&
  {Kroupa}}]{pflamm}
{Pflamm-Altenburg}, J., {Weidner}, C., \& {Kroupa}, P. 2007, \apj, 671, 1550.
  \eprint{0705.3177}

\bibitem[{{Pflamm-Altenburg} et~al.(2009){Pflamm-Altenburg}, {Weidner}, \&
  {Kroupa}}]{jan}
--- 2009, \mnras, 395, 394. \eprint{0901.4335}

\bibitem[{{Vanbeveren} et~al.(2007){Vanbeveren}, {Van Bever}, \&
  {Belkus}}]{vanbev}
{Vanbeveren}, D., {Van Bever}, J., \& {Belkus}, H. 2007, \apjl, 662, L107.
  \eprint{arXiv:astro-ph/0703796}

\end{thebibliography}

\end{document}